\documentclass[10pt,sigconf]{acmart}

\usepackage{booktabs} % For formal tables
\usepackage{indentfirst}
\setcopyright{none}

\renewcommand\footnotetextcopyrightpermission[1]{} % removes footnote with conference information in first column
\begin{document}
\title{A Novel Approach for Security Situational\\ Awareness in the Internet of Things }

\author{Fannv He}
\affiliation{%
\institution{National Computer Network Intrusion Protection Center\\
University of Chinese Academy of Sciences}
\city{Beijing} 
\state{China} 
\postcode{10000}
}
\email{hefn@nipc.org.cn}

\author{Yuqing Zhang}
\affiliation{%
\institution{National Computer Network Intrusion Protection Center\\
University of Chinese Academy of Sciences}
\city{Beijing} 
\state{China} 
\postcode{10000}
}
\email{zhangyq@nipc.org.cn}

\author{Huizheng Liu}
\affiliation{%
\institution{National Computer Network Intrusion Protection Center\\
University of Chinese Academy of Sciences}
\city{Beijing} 
\state{China} 
\postcode{10000}
}
\email{liuhz@nipc.org.cn}
 
% The default list of authors is too long for headers}

\begin{abstract}
Internet of Things (IoT) is characterized by various of heterogeneous devices and facing numerous threats. Modeling security of IoT is still a certain challenge. This paper defines a Stochastic Colored Petri Net (SCPN) for IoT-based smart environment and then proposes a Markov Game model for security situational awareness (SSA) in the defined SCPN. All possible attack paths are computed by the SCPN, and antagonistic behavior of both attackers and defenders are taken into consideration dynamically according to Game Theory. Two attack scenarios in smart home environment are taken into consideration to demonstrate the effectiveness of the proposed model. The proposed model can form a macroscopic trend curve of security situation. Analysis of the results shows the capabilities of the proposed model in finding vulnerable devices and potential attack paths, and even mitigating the impact of attacks. To our knowledge, this is the first attempt to establish a dynamic SSA model for a complex IoT-based smart environment.
\end{abstract}

%
% The code below should be generated by the tool at
% http://dl.acm.org/ccs.cfm
% Please copy and paste the code instead of the example below. 
%

\keywords{IoT, security situation, SCPN, Markov Game Theory}

\maketitle

\section{Introduction}

Internet of Things (IoT) defines a communication network that consist of highly interconnected heterogeneous devices [1]. IoT comes from the expansion of the Internet, which is destined to inherit most the security issues of the Internet. At the same time, new security issues occur to the IoT due to the numerous devices act as sensors [2]. IoT is still in the stage of development with no uniform standards for the hardware, software and communication protocols [3]. A variety of smart scenarios is also heterogeneous, which makes the defense of IoT extremely complicated. In order to perform a more effective and active defense as well as making defense decisions quickly when attacks happen, we are in expectation of real-time SSA of IoT-based environment.

Compared with the traditional SSA in the Internet, modeling SSA in IoT confronts the following challenges: 
(1) The SSA models in the Internet rely on the static network topology. However, the devices in IoT are ambulatory, which tend to a dynamic topology. (2) A SSA model in the Internet always has a unified communication protocol while various communication protocols exist in IoT system simultaneously and the devices communicate with others through their own protocols. (3) The SSA models in the Internet fail to consider the susceptibility of devices such as battery power consumption, while in IoT system the susceptibility should be stressed because some passive attacks caused by low power consumption often occur such as automatic sleep. 

Due to the challenges mentioned above, using the existing cyber security models [4] to achieve the purpose of real-time SSA in IoT is impossible. 
Several papers focus on developing security model for the IoT. 
Some of these papers [5-7] consider the vulnerabilities existence in IoT devices and propose a framework based on the network topology, but ignore the interaction between attackers and defenders caused by the offensive and defensive strategies changes in the process. Others [8,9] construct game-based security models for the IoT. However, their scope always address on some related issues in specific fields such as power consumption and consumer choice. Previous researches do not form a global, systematic and complete model for SSA, but instead mainly considers the security of IoT from certain domains. 

This paper defines a SCPN for IoT-based smart environment by extending the basic Petri Net. Then proposes a Markov Game model for SSA in the defined SCPN. Smart home environment is used to test the proposed model and two attack scenarios in the smart home are taken into consideration. The proposed model can simulate the curve of overall security situation. From the comparison of the curves under the two attack scenarios, we can find vulnerable devices and the potential attack paths related to them and mitigate the impact of attacks through modifying defense strategy.

The contributions of this paper are summarized as follows:

- We define a SCPN for IoT-based smart environment where colored tokens represent different types of threats. Threats propagation between IoT devices are captured by transitions, making the attack scenarios better exhibited, especially for collaborative attacks caused by various threats.

- We considered the interaction of strategy changes between the attackers and defenders in the attack-defense process, which makes the proposed model more applicable to the real smart environment.

- To the best of our knowledge, this is the first attempt to use Markov Game Theory in IoT-based SCPN to establish a SSA model for a complex smart environment.

The rest of the paper is organized as follows. Section 2 introduces methodology to develop the model. Section 3 describes a smart home environment with two attack scenarios in detail, provides experimental results and evaluates the performance of our model. Section 4 discusses some relevant issues of our approach. Section 5 concludes the paper.

\section{Methodology}
This section provides a methodological and theoretical basis for the concepts used in the paper.

\subsection{Stochastic Colored Petri Nets}
Stochastic Petri Nets (SPN) is suitable for modeling the dynamic behavior of any complex system [10]. CPN has colored tokens, which can describe various types of data and operations [11]. In IoT-based smart environment, attackers usually use synergistic attacks to achieve attack goals, which makes various threats propagate in the IoT. In this paper, we add colored tokens (represent different threats) to the SPN to expand it to a SCPN. Figure 1 shows a typical SCPN.

\begin{figure}
\includegraphics[height=1in, width=2.2in]{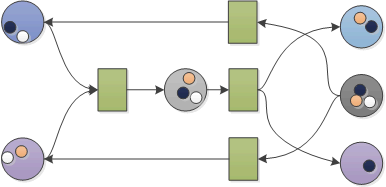}
\caption{A typical Stochastic Colored Petri Net, which comprises of places (circles), transitions (rectangles), directed connections and colored tokens (dots).}
\end{figure}
SCPN is defined as 6-tuple:
\begin{displaymath}
SCPN = \left( {A,T,P,C,\Im ,TR} \right)
\end{displaymath}

The components of SCPN are:

-Assets ($A$) are valuable resources in the smart environment, such as IoT devices and routers. Several vulnerabilities may be in these assets.

-Colored Tokens ($T$) represent different types of threats. There may be multiple tokens in one place. More tokens indicate that the corresponding node has been subjected to a more serious attack, and the node is more likely to affect other nodes. ${T^*}$ denotes one of the threats in the smart environment.

-Places ($P$) are all the possible locations of IoT nodes in the net. IoT nodes affected by a threat ${T^*}$ include nodes that have been attacked or may be attacked, expressed as $Nod{e_{{T^*}}} = \left( {P_{{T^*}}^1,...,P_{{T^*}}^i} \right)$.

-Directed Connections ($C$) are located between places and transitions, indicating the directions of threat propagation. The propagation path of a threat ${T^*}$ is expressed as $Pat{h_{{T^*}}} = \left( {C_{{T^*}}^1,...,C_{{T^*}}^j} \right)$. 

-Threat subnet ($\Im $) contains all the affected nodes and propagation paths associated with these nodes. The threat subnet of ${T^*}$ is expressed as ${\Im _{{T^*}}} = \left( {Nod{e_{{T^*}}},Pat{h_{{T^*}}}} \right)$.

-Transitions ($TR$) indicate the propagation of threats. When a transition occurs, the token moves from a node to another. $T{R^\xi }$ denotes the probability of threats propagating between two nodes successfully.
\subsection{Markov Game Model}
Game Theory (GT) captures the nature of conflicts: determination of the attacking-force strategies is tightly coupled to determination of the defense-force strategies [13]. Markov Decision Process (MDP) refers to the process that a decision-maker selects the behavior from the available behavior set based on the present state at each moment [14,15]. Markov Game is a combination of GT and MDP. The state at the next moment is only relevant to the current moment, so the threat transition has Markov properties. We establish a Markov Game model in the defined SCPN.

The proposed model is defined as 5-tuple:

\begin{displaymath}
MGM = \left( {\lambda ,S,\Omega ,\Phi ,\Re } \right)
\end{displaymath}

The components of the model are:

- Players ($\lambda $) set contains an attacker and a defender in this game. The attacker (${\lambda _A}$) spreads ${T^*}$ for damaging the performance of the system and the defender (${\lambda _D}$) cuts off the propagation paths of ${T^*}$ to keep the whole system stable.

- State Space ($S$) consists of all the possible states of ${\Im _{{T^*}}}$. The system states are determined by previous states and the current actions. The state of the ${i^{th}}$ node at time $\tau $ is ${\mu ^i}\left( \tau \right)$. The state of the ${j^{th}}$ propagation path at time $\tau $ is ${\eta ^j}\left( \tau \right)$. The overall state of ${T^*}$ at time $\tau $ is ${\Im _{{T^*}}}\left( \tau \right) = \left\{ {{\mu ^i}\left( \tau \right),{\eta ^j}\left( \tau \right)} \right\}$.

- Action Space ($\Omega $) includes all the possible combinations of actions in the game. At every time step, each player chooses strategies with associated actions. The attacker's action (${\Omega ^t}$) is propagating threat with a certain probability. The defender's action (${\Omega ^v}$) is to perform the strategy such as fixing a vulnerability, cutting off a propagation path, or removing a IoT node.

-Transition Rules ($\Phi $). The purpose of $\Phi $ is to calculate the probability distribution over the state space ($S$). 
The variation of the state is described by:

$\Phi \left( {{\Im _{{T^*}}}\left( {\tau + 1} \right){\rm{|}}{\Im _{{T^*}}}\left( \tau \right),{\Omega ^t}\left( \tau \right),{\Omega ^v}\left( \tau \right)} \right)$, 

where ${\Omega ^t}\left( \tau \right)$ denotes the action of the attacker at time $\tau $, ${\Omega ^v}\left( \tau \right)$ denotes the action of the defender at time $\tau $.

-Reward function ($\Re $). Since the purpose of the attacker is to maximize damage of the SCPN, its reward function is in relation to the damage. The purpose of the defender is to minimize the damage of the SCPN, its reward function is in relation to the reduced damage.

\subsection{Game Process}
Game Process involves how players make decisions under the interaction of each other and selects strategies from $\Omega $ according to the current state, and then get the one-step reward.

For the attacker, ${T^*}$ damages node $i$ and its associated propagation paths. The one-step reward function at time $\tau $ is expressed as:

\begin{equation}
\Re _{{T^ * }}^t\left( {{\Im _{{T^*}}}\left( \tau \right)} \right) = \sum\limits_{i \in o} {{\Re ^t}\left( {{\mu ^i}\left( \tau \right)} \right)} + \sum\limits_{j \in pat{h_i}} {{\Re ^t}\left( {{\eta ^j}\left( \tau \right)} \right)} 
\end{equation}

where, $i \in o$ is the nodes affected by ${T^*}$, $j \in pat{h_i}$ is the propagation paths related to node $i$. ${\Re ^t}\left( {{\mu ^i}\left( \tau \right)} \right)$ denotes the damage of the node $i$, $\sum\limits_{j \in pat{h_i}} {{\Re ^t}\left( {{\eta ^j}\left( \tau \right)} \right)} $ denotes the damage of the propagation paths related to node $i$.

For the defender, taking security strategy will bring two effects: reducing the damage produced by ${T^*}$ and affecting network performance inevitably. ${\Re ^v}\left( {{\mu ^i}\left( \tau \right)} \right)$ denotes the variation of the node $i$ after taking security strategy. $\sum\limits_{j \in pat{h_i}} {{\Re ^v}\left( {{\eta ^j}\left( \tau \right)} \right)} $ denotes the variation of paths related node $i$. The one-step reward function at time $\tau $ is expressed as:
\begin{equation}
\begin{split}
\Re _{{T^ * }}^v\left( {{\Im _{{T^*}}}\left( \tau \right)} \right) = - {\Re ^t}\left( {{\mu ^i}\left( \tau \right)} \right) - \sum\limits_{j \in pat{h_i}} {{\Re ^t}\left( {{\eta ^j}\left( \tau \right)} \right)}\\
+ {\Re ^v}\left( {{\mu ^i}\left( \tau \right)} \right) + \sum\limits_{j \in pat{h_i}} {{\Re ^v}\left( {{\eta ^j}\left( \tau \right)} \right)}
\end{split} 
\end{equation}

The threat propagates to the uninfected node through the ${\Im _{{T^*}}}$, the reward function is expressed as:

\begin{equation}
\begin{split}
{\Re _{{T^ * }}}\left( {{\Im _{{T^*}}}\left( \tau \right)} \right) = \Re _{{T^ * }}^t\left( {{\Im _{{T^*}}}\left( \tau \right)} \right) + \Re _{{T^ * }}^v\left( {{\Im _{{T^*}}}\left( \tau \right)} \right) \\
+ \omega \sum\limits_{\tau + 1} {\Phi \left( {{\Im _{{T^*}}}\left( {\tau + 1} \right){\rm{|}}{\Im _{{T^*}}}\left( \tau \right),{\Omega ^t}\left( \tau \right),{\Omega ^v}\left( \tau \right)} \right)} \cdot \Re _{{T^ * }} \left( {{\Im _{{T^*}}}\left( {\tau + 1} \right)} \right)
\end{split} 
\end{equation}
where $\omega $ is discount factor.

\subsection{Security Situational Awareness}
The proposed model considers the worst situation and evaluates the maximum damage of the system. The goal of the defense strategy is to maximize the defender's reward function for the maximum damage. The reward function for ${T^*}$ can be expressed as:
\begin{equation}
{\Re _{{T^ * }}}\left( \tau \right) = {\max _{{\Omega ^t},{\Omega ^v}}}\left\{ {{\Re _{{T^ * }}}\left( {{\Im _{{T^*}}}\left( \tau \right)} \right)} \right\}
\end{equation}

The security situational situation of IoT system at time $\tau $ can be obtained by summing up the reward functions of all the threats, which can expressed as:

\begin{equation}
\Re \left( \tau \right) = {\log _B}\sum\limits_{T \in Therats} {{B^{{\Re _{{T^*}}}\left( \tau \right)}}} 
\end{equation} 

where, $B$ is the radix.

\section{Experiment and Results}
The IoT has been widely applied in various fields [16-19], including smart home, healthcare, transport, environment monitoring, etc. We use smart home environment as a case to test the proposed model for SSA experimentally. We divide the devices into different regional subnets according to the communication protocols and model the heterogeneous subnets independently, then integrate these submodules together.
\subsection{Network description}
Fig. 2 shows the IoT-enabled smart home environment, which includes a ZigBee subnet and a Wi-Fi subnet. A smart home hub is used to support the communications of Wi-Fi, ZigBee and Internet, which provides users a control panel to access IoT devices and control them remotely. The smart hub establishes a ZigBee subnet that allows home devices (such as smart meters, thermostats, temperature) to communicate with each other by using the ZigBee wireless protocol. Android tablet connects to both Wi-Fi network and ZigBee network. Smart TV connects to the Wi-Fi subnet. Both tablet and TV have access to the Internet through the smart hub.

\begin{figure}
\includegraphics[width=3in,height=2.5in]{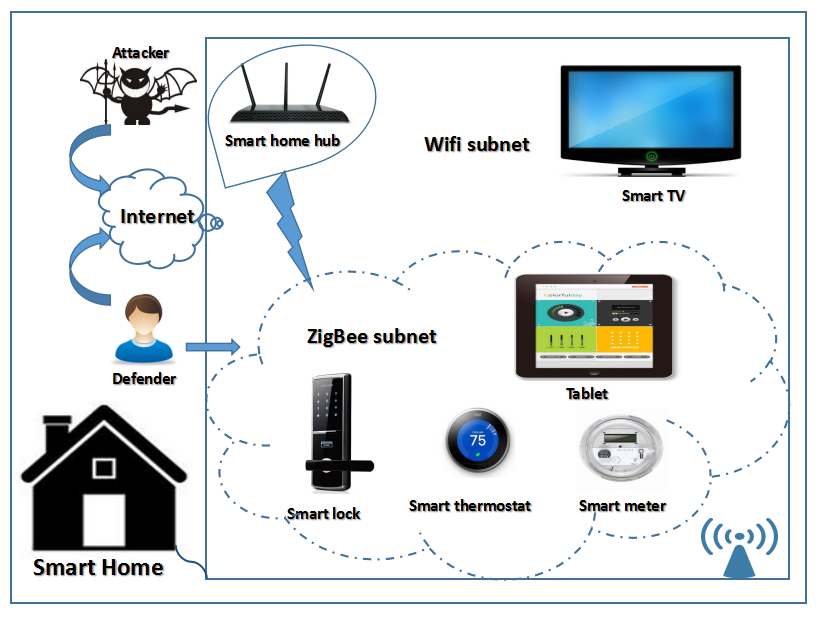}
\caption{A smart home environment.}
\end{figure}
\subsection{Attack scenario}
We assume the ultimate goal of an attacker is to damage the smart lock. Because the lock is isolated, an attacker cannot access it directly and only tries to perform infiltration attacks from other home devices. Two attack scenarios are taken into consideration in the smart home.

Scenario 1: We take the TV as a gateway to attack the smart home. Assuming that TV uses FFmpeg5.0. Attackers can exploit two types of vulnerabilities [20,21] in media file formats supported by FFmpeg5.0 to run attack code and gain the root privilege of the TV. Table 1 shows the information about the two vulnerabilities in the Common Vulnerabilities and Exposures (CVE) and their CVSS base scores. After getting root privileges, they can use the TV as a portal to exploit vulnerabilities in other devices such as Android Tablet. 

\begin{table}
\caption{Vulnerability information in the TV.}
\label{tab:freq}
\begin{tabular}{cccl}
\toprule
Vul&CVE ID&CVSS Base Score & Impact\\
\midrule
${V_{t{v_1}}}$&CVE-2008-4866 &10&10\\
${V_{t{v_2}}}$&CVE-2009-0385& 9.3&10\\
\bottomrule
\end{tabular}
\end{table}

Scenario 2: We take Android Tablet as a portal to attack the smart home. Assuming that attackers can write a malware to get the root permission of Android Tablet through utilizing three bugs [22] in the software and operating system. Table 2 shows the information about the three vulnerabilities of the Android Tablet. After getting root privileges, they use the tablet to launch other attacks targeting the ZigBee subnet.

\begin{table}
\caption{Vulnerability information in the Android Tablet.}
\label{tab:freq}
\begin{tabular}{ccc}
\toprule
Vul&Operation&Impact \\
\midrule
${V_{ta{b_1}}}$& Disrupt the conversion of Java bytecode&2\\
${V_{ta{b_2}}}$& Modify the AndroidManifest.xml file&2\\
${V_{ta{b_3}}}$& Obtain extended Device privileges&10\\
${V_{ta{b_4}}}$& Automatic sleep caused by low power&5\\
\bottomrule
\end{tabular}
\end{table}

\subsection{Experimental results and Analysis}
We create an asset list for the smart home environment to construct the SCPN, and establish the Markov Game Model in SCPN, then evaluate the security situational of the smart home environment under the two attack scenarios mentioned in section 3.2, respectively. Table 3 shows the states of nodes at time $\tau $ for ${T^*}$. AssLevel denotes the importance of a asset in the SCPN and have 5 levels. ThrOR denotes whether the node is infected by ${T^*}$. VulOR denotes whether a vulnerability exists in the node that can be exploited by ${T^*}$. Specifically, $\left( {{N_3},5,1,1} \right)$ denotes that the Tablet's value level is very high and vulnerabilities exist in the Tablet and can be exploited by ${T^*}$.

Table 4 shows the state of propagation paths at time $\tau $ for ${T^*}$. PathLevel denotes the importance of a path in the SCPN and also have 5 levels. Exploitability denotes the probability of a threat transmission through the path. Specifically, $\left( {{N_2},{N_1},5,3} \right)$ denotes that the propagation path between smart hub and Tablet is very important and utilizability of the path is medium.
Fig.3 shows the parts of constructed SCPN in the smart environment for ${T^*}$.

In order to facilitate the analysis of security situational value, we normalize the security situational values according to the min-max standardization. The results generated by the proposed model in the two attack scenarios are shown in Fig.4. The curves reveal the security situation trend overall, which provide a macro perspective of security situation in the smart home environment. 
\begin{table}
\caption{States of nodes under threat propagation at time $\tau $. ( for ${T^*}$)}
\label{tab:freq}
\begin{tabular}{ccccc}
\toprule
Asset&NodeID&AssetLevel&ThrOR&VulOR\\
\midrule
Smart hub & ${N_1}$ &5&0&1\\
TV&${N_2}$&4&1&1\\
Tablet&${N_3}$&5&1&1\\
Meter&${N_4}$&3&0&0\\
Themostat &${N_5}$&2&0&1\\
Lock &${N_6}$&5&0&1\\
\bottomrule
\end{tabular}
\end{table}

\begin{table}
\caption{States of nodes under threat propagation at time $\tau $. ( for ${T^*}$)}
\label{tab:freq}
\begin{tabular}{cccc}
\toprule
SouNodeID&TarNodeID&PathVlue&Exploitability\\
\midrule
${N_2}$&${N_1}$&5&3\\
${N_2}$&${N_3}$&5&4\\
${N_3}$&${N_5}$&3&1\\
${N_3}$&${N_6}$&1&1\\
\bottomrule
\end{tabular}
\end{table}
\begin{figure}
\includegraphics[height=1.8in, width=2in]{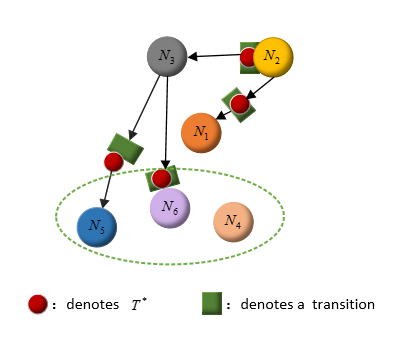}
\caption{Parts of constructed SCPN in the smart environment. ( for ${T^*}$)}
\end{figure}

\begin{figure}
\includegraphics[height=1.5in, width=3in]{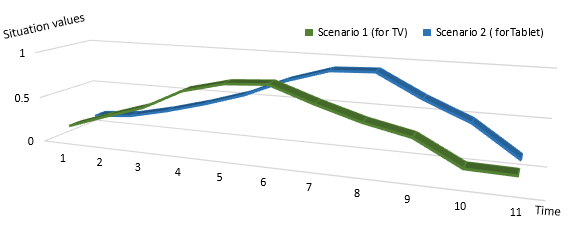}
\caption{Security situational curves of two scenarios.}
\end{figure}

Furthermore, according to the comparison in Fig.4, we can find that attacking the TV has faster attack effect but lower impact than attacking the tablet. Thus, the attacker is more likely to choose the TV as the entry point. The defender should protect the TV at first in order to prevent the attacker from breaking into the network. Moreover, they can take TV as an origin and find the potential attack paths from the SCPN, then decide which IoT devices included in the paths should be protected at the same time.

\section{Discussion}
In this section, we clarify some relevant issues of our approach and discuss the limitation of this approach. 

In this paper, we establish a SSA model in a complex IoT-based smart environment. The proposed model can form a macroscopic trend curve of security situation and help administrators make effective defense decisions to mitigate the impact of potential attacks. 

Compared with the previous models,
(1) We chose SCPN to describe the IoT, because SCPN will be created dynamically which will make IoT-based smart environments highly interoperable and scalable, and provide IoT system a dynamical topology.
(2) We divide the IoT devices into different regional subnets according to the communication protocols and model the heterogeneous subnets independently, then integrate these submodules together thus solving the problem about heterogeneity of IoT devices.
(3) We stress the susceptibility of devices in this approach, for example, passive attacks caused by low power consumption are taken into consideration.

It is worth mentioning that although we validate the proposed model via two attack scenarios in the smart home environment, our approach is portable that applies to other IoT environments through modifying the parameters of the model.

On the other hand, our approach considers the behaviors of both player and the dual effects of propagation nodes and propagation paths in the attack-defense process, so the state space would be large when players making decisions. Especially in complex IoT environment with numerous IoT nodes and propagation paths, the resources consume is huge. Hence, we will focus on reducing resources consume through combining the IoT vision with cloud computing.
\section{Conclusion}
Internet of Things (IoT) is enabling innovative applications in various domains. Due to its heterogeneous and wide-scale structure, it brings many new security issues. In this paper, a SCPN is constructed for a IoT-based smart home environment, and a Markov Game model is proposed for SSA in the defined SCPN. All possible attack paths are computed by the SCPN, and antagonistic behavior of both attackers and defenders are dynamically taken into consideration according to Markov Game Theory. We evaluated the proposed model in two attack scenarios in a smart home environment. The proposed model can form a macroscopic trend curve of security situation. According to the analysis of the results, we can find the vulnerable devices and the potential attack paths related to them in the SCPN, and then choose effective strategies to protect the devices and mitigate the impact of potential attacks.

\section*{Acknowledgment}
This work is supported in part by National Key R\&D Program of China (2016YFB0800700), the National Natural Science Foundation of China (61272481, 61303239,61572460),  the National Information
Security Special Projects of National Development and
Reform Commission of China [(2012)1424], open Project
Program of the State Key Laboratory of Information
Security(2016-MS-02).

\end{document}